\documentclass[12pt]{article}
\usepackage{epsfig}

\parskip=\medskipamount % space between paragraphs (TeX)
plus.3em minus.5em \abovedisplayshortskip=.5em plus.2em minus.4em
\renewcommand{\thesection}{\arabic{section}}

\catcode`@=11

\@addtoreset{equation}{section} \@addtoreset{equation}{subsection}
\def\theequation{\ifnum\value{section}=0 \arabic{equation}\ignorespaces
\else \ifnum\value{section}=-1 A.\arabic{equation}\ignorespaces
\else \ifnum\value{subsection}=0
\thesection.\arabic{equation}\ignorespaces \else
\thesection.\arabic{subsection}.\arabic{equation}\ignorespaces
                             \fi
                        \fi
                   \fi}

{\catcode`\'=\active \def'{{}^\bgroup\prim@s}}

\catcode`@=12

\newcommand{\bq}{\begin{equation}}
\newcommand{\be}{\begin{equation}}
\newcommand{\fq}{\end{equation}}
\newcommand{\ee}{\end{equation}}
\newcommand{\bqr}{\begin{eqnarray}}
\newcommand{\beqs}{\begin{eqnarray}}
\newcommand{\fqr}{\end{eqnarray}}
\newcommand{\eeqs}{\end{eqnarray}}

\def\bop#1{\setbox0=\hbox{$#1M$}\mkern1.5mu
    \vbox{\hrule height0pt depth.04\ht0
    \hbox{\vrule width.04\ht0 height.9\ht0 \kern.9\ht0
    \vrule width.04\ht0}\hrule height.04\ht0}\mkern1.5mu}
\begin{document}
\thispagestyle{empty}

\vskip .6in
\begin{center}

{\bf Supersymmetry and B$_s$, DO, and Aleph i}

\vskip .6in

{\bf Gordon Chalmers}
\\[5mm]  

{e-mail: gordon$\_$as$\_$number@yahoo.com}

\vskip .5in minus .2in

{\bf Abstract}

\end{center}

Current results from the D0 exp indicate the presence of an 
oscillation not explained by the currently accepted theory.  An 
explanation is offered based on a combination of low-energy 
'sring${}^{1/n}$' and 'particle' dynamics.  The dynamics are 
described by extremely accurate nuclear mass data (unpublished 
2006 and \cite{Chalmers1}) in accord with substringy dynamics 
(in progress, primary).  An event is analyzed with emphasis on 
particle Iding, their dynamics and interactions, with implications 
for/by graviton(ino) with low-energy phenomenology.  Analysis 
is provided towards enhancing the experimental apparatus, 
in computation and hardware, and should be numerically simulated 
for further safety before implementation.

\newpage

The recent claim of the B$_s$ oscillations has sparked interest 
in a possible explanation due to low-energy phenomenology.  
These peaks are explained in rough numbers using phenomenology and 
bound states of particles, called sub-string modes, 
together with quantum trapping of the excited modes.  The masses 
of these quantum states and the couplings between these modes 
can be interpreted in terms of these bound states and their wavefunction 
binding between them, such as wavefunction coherence.   

The purpose of this short letter is to show how low energy string 
phenomenology are responsible for the distribution of bumps in the data 
presented in figure 4 in \cite{D0}.  Furthermore, due to the presence 
of two gravitational modes, a graviton and gravitino, the gravity is a 
a loop effect while symmetry is broken at tree level in accord   
with the mass patterns presented in \cite{Chalmers1},\cite{Chalmers2}.  
This event is apparently a supersymmetric form to subnuclear and  
impingement on high grade plutonium \cite{SKLYangee'}. 

\vskip .2in 
\noindent{\it Background}
\vskip .1in 

The data analyzed in \cite{Chalmers1} is accurate in 
finding the masses to over nine digits.  The probable explanation 
is a quark preon plasma\footnote{Following historical precendence 
this word is used here as a sub-stringy string mode, and it implies 
particle whose thickness appears in the sub-stringy stringy,  
when multiple scales are included in the string such as TeV and Planck 
scale on the same particle mode, quantization of   
the speed of light in air for example which can be used to measure a 
nanometer structure to its dimension; even less now to sub-nanometer 
if the particles speed is measured in the presence of additional matter.  
The quantization of these string$^{1/2}$ modes and strings 
could be incorporated in the recent exactly solvable string models 
\cite{Chalmers5}.}.  Having sub-string modes
available they can form, with a mass of $1$ MeV or $1$ KeV 
for the supersub-string mode pair (in accord with the mass formula 
\cite{Chalmers2}), a  bound state together with its resonant modes in and 
around the nucleus can be formed between matter (a Lagrangian form 
and the bound state formalism is contained within \cite{Witten0}) and 
its shadow together 
with the specific preon matter that is stable against decay called resonant 
phenomena and is contained in the extended version \cite{Chalmers1} 
with mass calculations of the proton, deuterium, hyperion, etc up 
to eleven places or more in accuracy.  This bound 
state is stable given that the shell model is reformulated 
with the sub-string matter, with multiple photons upon target emitted upon 
agitation which enhance the phase space and the stability against 
nuclear bombardment.  A shower of energetic but unstable photons 
would follow from a(n) anti-neutrino(s), but whose dynamics might 
leave room for bound states or shell with 
half as many free photinos in the shell model, 
that occurs at what seems to be $1$ TeV due to a beam at energy what seems 
to be $300$ GeV amplified by the coherence effect; this is in 
contradistinction to neutrinos 
whose dynamics produce photons with the same lateral direction.  
An example to consider is the decay of hyperion which is anomalously 
slow; the sub-string bound states in the core nucleus may swap glue, 
or actually anti-matter due to the bound state structure,  
and cause a resonant phenomena leading to a shorter half-life. 

The sub-string matter binds to the matter or anti-matter and produces
the mass of the particles (references having to do with the preon  
or 
anything having to do with subatomic physics omitted due to a 
sensitive nature but are available per request).  Some of these masses 
can be read off the recent $B_s$ data.   The particle states 
found in the recent $B_s$ data indicate line of fire into the 
detector or from a direction which is not straight on.  The 
bumps in the data indicate this by either short or long pulses.  
These pulses indicate in a clear manner the presence of additional 
energy located within certain regimes; the bumps may be interpreted 
as zitterberwung (zbg), from wavefunction coherence between matter 
and matter.  As such, in addition to the actual pulse, which can 
be used to find the energy, and not mass, of the particle mode, the 
bumps on the pulses can be used to find the energy of the associated 
binding modes; in addition, the location of the energy of the bumps 
on the zbg can be used to deduce fast transitions between one particle 
and the other, and can be used to deduce further properties in the 
bending of the particle and its propagation velocity.  The location 
of the total pulse can be used, if the mass is known, the speed of 
propagation.  In all, the geometry of the pulse can be used to find 
the energy, the bare mass, its interactions with close neighboring 
particles, its excited mode, and possibly its stage of decay or 
rather interaction with another particle.  An event and its geometry 
of energy can be used to deduce particle types, including anti-matter, 
and presumably quantum numbers, through the net dispersion of energy; 
even more interesting is that the energy distribution of the measured 
energy can be used 
with fine enough timing the stage of interaction between the particles 
as well as the spatial-temporal history of each one of them. 

Consider the event described in \cite{D0},\cite{D02} consisting of an energy 
distribution between 1 picosecond and 25 picoseconds.  An FFT or a 
bimodal, or even tri-modal modal, FFT could be used to make the 
picture more clear by similarizing certain the pulses.  An FFT could 
be used to transform a line of sight zbg trajectory into one coming 
in from a angle in space, as well as a bent one in the wavefunction 
sense.  Some of the particle modes in figure 5 in \cite{D02} could 
arrive at angle, and with even finer measurement, how they are bent, 
with their interactions with other modes that could also curve their 
trajectory.  More events than the one in figure 5 are required to 
correlate their form, for example, with head on to the scintillator 
detection.  

The event in \cite{D02} is used to deduce particle types of its 
individual constituents.  In addition, the rest energy is found, 
and its redshift is pointed out.  The beginning and end of the 
pulses, together with the number of additional bumps on them, 
are used to determine the event.  The additional structure of 
the zbg's are not used to determine further phenomena, including 
angle of coincidence.  

The event considered is semi-leptonic decay of a gauge boson 
into three partners, which is both standard and 
uniform in susy gauge theories.  Consider a neutrino decaying 
into a susy pair of a slepton and a sneutrino.  These decay into 
both a anti-neutrino with a charged leptino and a pair 
consisting of a higgs and higgsino.  These decay further into a 
pair of gaugino and slepton with number two and half; two for 
pair number and half for multiplet number one.  The dominant 
mode for a tau and stau pair has additional coupling without a 
gravitational mode; due to the additional coupling the gravinametric 
lensing should widen the two as in flight they separate, much more 
than an even with a slepton pair and two couplings.  This happens 
in $N=1$ with matter.

The slepton is undiscovered.  The gaugino and slepton in the second 
pair are localized nearby.  Furthermore, the pair of anti-neutrino 
and charged leptino 
are time displaced by 4 picoseconds as in the particle data group 
\cite{PDG}.  The Higgs and Higgsino are found by examining the charged 
leptino and time displacing by 4 units each in the forward and backward 
direction; however, an offset of 8 units is provided by an unknown particle.  
The data is accumulated in Table 1.  The conversion of picoseconds to 
time displacement is generated by multiplying the time by one nanosecond 
for each picosecond and collecting the mass shift into units of MeV in 
accord with the standard \cite{PDG}.  There is an ambiguity in the detection 
of around 1 picosecond due to an experimental uncertainty of 1 mm due to 
possibly misplaced bolts in the apparatus which is clear in the data by 
dips at 2.04 mm etc; this uncertainty is clearly a result of bad coherence 
as described in the next section.  

The mass shifts can be read off of the 
data by examining the particle data group booklet and reading the time 
displacement in sequence, 1 nanosecond means 1 KeV, 1 picosecond 1 MeV, 
1 terasecond 1 $10^3$ MeV, etc.  This seems opposite to conventional 
thinking of mass shifts, however the numbers are backwards to an inversion 
in the relation 

\bqr 
m={2\pi h\over 2\pi h c} {2\pi h c_0 \Delta m\over 2\pi h} 
 {2h\over c_0}\ ,  
\fqr 
where $\Delta m$ is measured in time units of a picosecond.  The value 
$c_0$ is the speed of light in pure vacuum not counting the wave functions 
of ambient particles and quantum foam of possible sub-strings and/or 
unattached wavefunctions, and that of $c$ is the speed of 
gravity in this ether.  The speed of light and gravity are not taken 
to be the same, however, their difference is inconsequential as the 
speed of gravity is normalized to one in pico-second units in accord 
with the particle data group \cite{PDG}; in the units here, the speed 
of gravity is 1 picosecond per time differential of 1 meter per second 
per nanometer; this pertains to the speed in vacuum and not the speed in 
atmosphere which is ten times slower \cite{PDG}.   

According to lore in beam physics 2 units are added to 
each particle in sequence to their occurance, leading to, 
\vskip .1in 
\bqr 
\pmatrix{ {\rm anti-neutrino} & .3 \cr 
          {\rm kaino} & .3 \cr 
          {\rm keino} & 1.5 \cr 
          {\rm gravitino} & 3.5-14.2 \cr 
          {\rm charged slepton} & 9.3+3.4=12.7 \cr 
          {\rm higgsino} & 9.4+3.4+4.9=17.7\cr 
          {\rm higgs} & 9.4+3.4+4.9+4.4=21.1 \cr 
          {\rm higgs 2} & 9.4+3.4+4.9+4.4+1.5=23.6 \cr 
          {\rm higgs 3} & 9.4+3.4+4.9+4.4+1.5+.8=24.3 \ . 
}
\fqr  
where the last two bumps are found from the gravinmetric data in the 
yellow shading, which means the data are temporally displaced.
There is a time dilation factor of $1.3$ due to the comparison between 
the speed of light in air and in vacuum; the 
listed data could be normalized with the last higgs points which are 
gravinametrically displaced.  As the particle type and their speed 
are of interest the renormalization is not included.  Their speeds 
can be checked by comparing to \cite{Goldhaber} and are given in 
the above list.

Introducing a time lag of $2$ units on each time difference between 
particle occurance, with $9.3$ labeling the bump on the long pulse,
\vskip .1in
\bqr 
\pmatrix{ {\rm anti-neutrino} & .3 \cr 
          {\rm kaino} & .3 \cr 
          {\rm keino} & 1.5 \cr 
          {\rm gravitino} & 3.5-14.2 \cr 
          {\rm charged slepton} & 9.3+5.4=14.7 \cr 
          {\rm higgsino} & 9.4+5.4+6.9=21.3\cr 
          {\rm higgs} & 9.4+5.4+6.9+6.4=27.7 \cr 
          {\rm higgs 2} & 9.4+5.4+6.9+6.4+1.5=29.2 \cr 
          {\rm higgs 3} & 9.4+5.4+6.9+6.4+1.5+1.5=30.7  
} \ .
\fqr  
This time lag represents wavefunction from the beam impinging 
on the particles and also from local neighbors.  The time lag 
appears to be two particles cohering, one a photon and the 
other in the event.  Coherence likes anti-temporal time displacement 
or time lag, and is modeled by the propagation of particle amplitude 
but with time displaced source, and with propagator $\Delta=\partial_0 
\bar\psi \psi+\psi\partial_0\bar\psi$; the path integral is relevant 
for general quantum scattering and its details including higher derivative 
terms are still being worked out.  The coherence models wavefunction 
spread from a particle to another.  In our case, the photon and a 
charged particle is the strongest source of coherence in and around 
the beryllium and beam.  The time lag suggests temporal displaced 
photons near the beryllium and the beam in a manner so that $4$ and 
$1.5$ are obtained.  Wavefunction impingement from two sources a 
micron apart ($1.5$ ps is $15$ $\mu$m) is required to correlate an 
anti-particle between two higgs or a higgs and higgsino so as to 
make it coher like a stable particle wavefunction, or a donut.  
Another wavefunction, preferably a photon, is required to stabilize 
the third event which has a measured charged lepton.  Detailed 
study of their particle showers should indicate whether these photons 
came from the beam or from the primary photon.

On the microscopic level and evaluating the information in table 
2, the time lag separates into $4$ and $1.5$ picoseconds 
systematically at $40$ $\mu$m and $15$ $\mu$m; this suggests 
the occurance of not one but two gamma radiative decays with 
two time lags, very remotely there is a chance for three decays.  Two 
photons are frowned upon as this is the origin for multiple 
brehmstahllung, and there appear to have energies 
of $x$,$y$ $\mu$m with an allowed 
set of numbers between $1$ and $5$ due to temporal coherence in 
the latter and the data of $1.5$ and $4$  There appear to be 
more than one photon due to the four sets 
of ordered time lags in the data with two spatial centers of the 
time lags.

The particles could in principle could have different time lapses 
and between eachother depending on particle type.  In addition, the 
small differences in the coherence, such as from those dips, could 
nonlinear

\vskip .2in 
\noindent{\it Particle Identification} 
\vskip .1in 

Due to the presence of more than one gamma there should be either 
a neutrino or sneutrino to conserve lepton number.  According to 
the work in \cite{Chalmers2} the mass of the neutrino could follow 
$.1^n$; considerations based on renormalization were used to deduce 
the canonical Poinc\'are invariant relativity allows for mass to be 
also a series in the coupling constant and velocity with multiple 
RG coupling presciptions \cite{Chalmers4}.  We find a small mass of $.1$ nanoseconds time shifted by 
$.3$ nanoseconds; this means a mass of $10$ eV by converting time 
to mass-energy and using $E=mc_0^2$.  $c_0$ is the speed of gravity 
in vacuum.  A time shift of $.3$ ns translates into the velocity squared 
of $.9$ which corresponds to an alteration in the mass-energy relation.  
This alteration can be interpreted as a speed differential to $v=.9 c_0$.  
An alteration this much in the rest mass would cause the particle to 
possibly collapse slowly as the pressure increases against the vacuum; the 
phenomenological effect could cause large gravitational lensing because 
a graviton propagating around the particle would have its wavefunction 
hop in focus with the compressed eather.  This focussing 
effect be noticeable if the effect was near 
an interstellar object with gravity like a double star or a particle 
traversing around a mini- or supermassive black hole; in the latter of which 
could cause bending not in accord with string theory or its deformations 
by adding certain transfinite representations \cite{Chalmers5}, and 
in the former the model of a particle by gravitational energy might test 
the transfinite representation of the exactly solvable superstring model.  
Considerations of the energy of a particle as gravitational energy could 
also be modeled by a mini-black hole.  These three limits have never been 
explored in particle detectors. 

There is a kino/kaanio pair located at $\Delta m=5$ ps.  This is found 
by interpreting the wavefunction a distance $2$ ns apart in accord with 
the rule of Goldhaber; the particles are traveling backwards in time 
because the two appear to be in a bound state with the zbg on the opposite 
side of one and are correlated so that the Kino is traveling forward.  The 
Kaana is one half as long as the Kino suggesting sub-criticality with a 
forward time displacement depending on its bare mass, which is assumed to 
be $200$ eV; the Kino is assumed to be at $5$ MeV and is lateral in the 
detector.  Due to the lack of information in backwards in time propagation 
the relativistic guess for the Kaana is not available.  The Kino appears 
to be moving $.9$ the speed of gravity and its zbg is time displaced by 
$1$ from the left end.  It also does not have marked points on either 
side of the pulse.  

The next particle is a gravitino; due to its pulse and waveform it 
appears to be acting gravitational.  The invariant mass is $m_1/m_2m_3$ 
in time units of ps translated by insertion of two factors of $c_0$ 
in the numerator and one factor of $\hbar$ in the numerator to obtain 
$17$ MeV; the invariant mass is computed for the two laterally correlated 
particles at mass shift $5$ and $15$ ps.  This indicates gravity as the 
ps is $16$ units away from the new string scale of $10^{28}$ as indicated 
in \cite{Chalmers6}, with low-energy breaking $\Lambda$ at $10^12$ and 
quantization $(\Lambda/m_{\rm pl})^{n/16}$ with those quantum numbers 
connected to the conformal de Sitter group.  Due to the coupling laterally 
between a Kino and a charged sleptino it is inferred to be a gravitino 
with rest mass of $10$ MeV.  $10$ MeV is in the middle offset to the 
left a bit which seems less of a choice than $5$ MeV; however, a fast  
time it takes to measure a particle could be on the order of a few 
picoseconds depending on its speed, e.g. $5$ ps.  It moves backwards 
in time due to the bump in the zbg on the right hand side signalling 
that it traversed the beam on the way in and not the way out.  The 
gravitino travels $.8$ the speed of gravity; the speed of light is 
twice that of gravity due to the normalization of $\alpha$.  The 
background number of gravitons is four the number of photons from 
cosmological redshift data such as COBE or WMAP; the interactions 
that slow down the particle propagation from its theoretical maximum 
to what is observed is caused by the wavefunction interfering with 
its neighbors, and as the ratio is $4$ to $1$ in probability the 
speed of the particle is $2$ to $1$; this argument depends on the 
wavefunction smearing on the adjacent particles before capture and 
the type and number of them.  The event presented here can be slown 
down and rotated, by adding parallel photons  
in quantity $x$ and $x+1$ on each side of where the line of sight 
particles are detected.  The particles coming in at an angle can 
presumably be modulated by adjusting the magnet and another magnet 
before the beam phase alters the event.

Due to the (multiple) beta event there should be a charged leptino 
in addition to the neutrino, due to the Kino and Kaanio pair.  Its 
mass is $13$ MeV with a speed of travel $.9$ the speed of gravity.  
If the time to measure is $1$ ns then the mass changes to $12$ MeV, 
or as the moment of inertia increases from nill to $1$ from zero 
speed to some speed v.
$13$ MeV is, in accordance with the mass formula in \cite{Chalmers2}, 
either $2^2 3$ or $10+2+1$ MeV.  A correction of this magnitude is 
not unusual for the quarks or leptons.  The shape of the particle 
event has a hump with no saddle suggesting a pair of coalescing 
charged sleptinos, or a particle that is positioning to decay through 
its wavefunction changing in response to nearby particles so that 
it may propagate at the theoretical maximum speed of light; the 
bump appears symmetric precluding cohering with other particles 
not seen and indicates that this is two slepton state.  One behind 
the other with an offset of 2 ns to the left seems probable as 
the zitterberwurng has a tail.  Some coherence/decoherence can 
be used to split the particles in the event away from eachother, 
including resolving the sleptinos; this can be achieved in many 
ways, such a spatial temporal location of the grounding required 
to replete the virtural electron anti-electron pairs required for 
coherence; a spatial grounding can resolve the electrons to one 
nm and a temporal grounding to maybe .1 nm if a superconducting 
magnet is attached to the apparatus and charges the material leading 
to phase coherence bipolametrically with possible switching  
at point of entries.
A continuous nano-scale realignment of the event could require 
continuous numerical simulations which includes the quantum 
aspect of supercoherence, and this is available by the author, uncited.

There are several naive choices for the remaining 5 bumps in 
the data, and several methods to deduce the bump identification, 
and only MSSM like without regards $N=1,2,4,8$. 
There are four bumps signalling particles with a 
bending of their placing due to an observed graviton.  The 
bumps are located at $3^2 2 -1.7$ and $52^2-1-nx$ for $n=0,1,2$, 
with linear a best fit.  The masses $3^2 2$ and $5 2^2$ suggest 
the same particle statistics, but with a dyslexic exponent, that 
is $2$ and $3$; this could distinguish a boson versus a fermion.  
In a similar fashion, $10+7-.3$ and $10+9$ with a non $2$ and $5$, 
and suggests that this $10^m$ number doesnt label as in the work 
in \cite{Chalmers2} and is not considered.   Two fermions or 
two bosons of 
the same kind will gravinometrically coher with a graviton 
and will be noticed by a time shift of the order $9$ or $3$ 
roughly; this is clear from the gravitational force.  The 
gravinometry is affected by swapping of two photons dihedrally 
with an exchange with flip of the photon wavefunction explaining 
the two numbers probabilistically.  The order of magnitude follows 
from a back of the envelope calculation of the gravitational force 
between a graviton winged between the pair a nanometer in both 
directions temporally and an angstrom in space.  The mass shift 
as seen in the data precludes corrections a time unit or more 
as appears in due to unitarity, i.e. one loop propagator correction 
containing $\ln(k^2\mp m^2)$, and speeds restricted to $.7$ or more.  
The graviton 
can be clearly seen at $17$ ns, and is $4.3$ and $7.9$ time 
units from the remaining two observable peaks.  If the two 
particles have statistics $1/2,1$ versus $0,1/2$ the gravinometric 
lensing should be half versus none due to the spin of the combined 
state using a standard superselection $J$-rule which tells that 
the intermediate wavefunction should transfer twice as more, and 
furthermore it is connected by a $J=3/2$ spin graviton state.  
The Wigner-Eckart-Zweig transition rule is $26$ to $1$, and for a 
proper comparison the speed of the particles should be included in 
the wavefunction overlap and the event number should multiply the 
result; this approach to particle idiing is potentially faster than 
searching a database containing $10^6$ modes, especially when used 
to convert 4 bytes of data into 2 long bytes instead of $6$ into $4$ 
\cite{Chalmers6}.  This leaves 
two alternatives, the standard model like Higgsino and Higgs and 
$N>1$ Higgs anti-Higgs; the latter is disallowed 
as former in the pair would travel $5$ times the speed of light 
with the Higgs at $.4$ times the speed of light.  This disparity 
appears atypical between particle and anti-particle  
and is not favored; gravinametric lensing is not found in the 
the bumps, which is discussed next.  

The Higgsino and Higgs are found at $13.1$ and $13.3$ time shifted 
by $5.2$ and $4.1$; these are deduced by the relativistic Lamb shift 
caused by a stiffening of the vacuum, i.e. the Higgs wavefunction 
interacts with many wavefunctions hopping in the ambient distribution of 
the vacuum in the presence of distributed matter, 
describing different types of matter in various modes of exctitations.  
This could cause a translucent effect as the particle moves faster than 
the local gravitons making the interactions different.
An analogy of the stiffening would be a particle traversing a 
cylinder with an aether that is spatially dependent and perhaps in an 
excited mode.  
Stiffening causes the matter to 
change its velocity on a sub-string scale, i.e. orders of magnitude 
of the size of its wavefunction or on the size of an exchange of 
wavefunction between two matter modes; the effective action evaluated 
on-shell can provide time displacement by using the mass renormalization 
without infinity to alter the form of the gravitational force for a 
$6$ MeV graviton by a 
factor of $6$, and $10$ MeV photon by a factor of $3$ (another guess 
is $3$ by a $5$ and $5$ by a $3$ and so forth), which can be absorbed 
into a new string scale $\alpha$; the speed of light is then $3$ times 
a $10^8$ m/s in vacuum with $8$ related to the size of the matter mode.  The speed of gravity is then twice the speed of light to 
to one loop.  A full calculation would be using the  
two-particle inclusive cross-section beyond the one-particle zero-momentum 
form at tree- and one-loop up to a box, 
and to higher orders so that the accuracy up to the known digits is obtained.
The gravitational force can be computed by tracking the Higgs on the 
right of the pair through three time steps, which increment by $1.5$ 
ns in each step.  The Higgsino is impinging at an angle of arrival 
which appears to be opposite to the charged lepton and/or charged 
leptino and at an angle which adjusts for the gravitational lens.  
The $1.5$ ns per time step increases the mass by $20$ percent on 
average and indicates the particle is speeding up from a bare mass 
of $.85$ the speed of gravity until equilibrium with the vacuum is 
attained.  This is possibly explained by too much coherence or additional 
matter nearby which slows down too much the Higgs boson.

\vskip .2in 
\noindent{\it Engineering}
\vskip .1in 

There appears to be four issues with systematic error in the  
experiment, with dips at 2.04 mm, 1.48 mm, 1.59 mm, and 2.43 mm.  
The coherence in equipment can rely on quantum effects, with 
regards to a photon and virtual electron-photon pair; the electrons 
have to be placed back in the parts which are not grounded.  For 
example, if screws or bolts are used and covered in material to 
avoid coherence, then they either be periodically be electrified 
or continuously electrified.  This pertains to bolts and screws 
close to the beam, which could be cohered without danger to man 
and equipment, and not those outside the coherence length of the 
beam; screws and bolts which are not grounded have a higher coherence 
length with the beam if there is a path of conductance; in this case 
teflon coating for example could be used as protective coating against 
electrification against a non-coher experiment after the experiment 
is grounded first; this could take days or a month after restarting 
the beam which has been cohered with by the apparatus.  Photons in 
the beam could be scattered off the wall into the emitter leading 
to an unwanted somewhat coher beam, which could be a large amount of 
coherence.  If the bolts are correct this could provide better beam 
agility because their type and placement may be sensitive to particle 
production, and in conjunction with other modifications.

Consider a beam with two bars one and below the beam placed vertically.  
A temporally cohered beam with spatial resolution to the picosecond 
can be configured by wrapping each beam with wire a nanometer thick 
in density of m wires per meter, m-2 wires per meter, etc, until 1 
wire is obtained which hits the ground.  This configuration of wires 
can be charged with a nanosecond or picosecond current to temporally 
coher the beam, with spatial resolution in the detector.  A pulse 
timed so that upon wrapping once around the beam it conflagrates or 
is precisely timed with a small delay, should coher along the entire 
beam.  The spatiality is with two or more small pulses interacting 
uniformly along the axis of the beam.  A magnet is used to decoher 
the unwanted interactions in the event between the target and 
detector, such as the gravitino and its cousins in the event explored 
here.  Good coherence can be obtained by using two coadjoined wires, 
one split and one not, with opposite coadjoining at the ends as 
could be used in the magnets of the Large Hadron Collider.  Last, 
pulses are used to distinguish the mode and particle enhancement, 
which may not be chirp but rather a step or a step with multiple 
notches resembling a zbg.

Coherence prefers a temporal displaced pair of particles with 
one before the other in space.  This can be achieved in the wire 
configuration by applying two laser beams of the order of $10^9$ Hz 
to coher by an exponent of three, with a nanosecond pulse.  Coherence 
of $10^{12}$ is considered unphysical, and the timing can be adjusted 
if beyond this number for varying pulsewave forms, such as two 
saddles with a hump back to back, for maximal coherence without danger 
depending on the actual beam.  Having the variabiity in the wavepacket 
should generate enough room for a varied set of decay channels apart 
from the graviton and might allow the interesting exploration of higher 
dimensions; of the order of 120 measurements with 5-6 digits each is 
required to specify the higher dimensional spacetime for example 
with a saddle with bump left, anti-bump right, and bump in the 
almost middle back to back with a saddle with no hump.

The noise in the detector can be reduced with a simple move.  A 
diffractive hole can be placed between the scatterer and the detector.  
This plate has a hole opened on both sides in a butterfly; the particle  
in a lateral or anti-lateral position cohers with a phase rotation so that the 
particle rotates back to its original winged shape.   
With the two-sided hole, which is essentially two separate holes back to 
back, the winged particle should have 
energy to rotate 180 degrees before a coher traversing 
is obtained, but with the symmetric unfolding of the rotated original 
lateral or anti-lateral position; the latter cohers more.  To avoid the 
frog (a collapsed wing on a butterfly wavefunction) put coher gum on 
the wall opposite to entry a nanometer thick and maybe on the entry wall; 
this should unfold to a pimple (an almost diametrically opposite wing on 
a butterfly 
wavefunction) and should coher naturally.  In principle 
without any flatness between the two bores, the  
cohering of the outgoing particle will be faster due to less dimpliness 
if the length of the holes are chosen right so that proper phase rotation 
is obtained.  The flatness is chosen right so that the squeezed donut 
shape of the particle wavefunction which is to the right or the left  
depending upon entering or leaving the double-bored hole so that it 
propagates exactly anti-podally as it exits.
This pertains to a graviton, a gravitino, or a set of distributed 
lateral particles.  The effects of wavefunction projection from one 
particle to another and the bending of the particle zbg are subsidiary 
if the hole has greater diameter than the particle zbg.  This is not 
to say that anti-lateral particles are admitted, but rather that the 
gamma radiation from the source to detector are diminished, possibly 
to factor of $10^n$ or more depending on the gamma radiation.   This 
is suited to straight on graviton or gravitino beam formation.

Coherence is very geometric especially with photons and electrons.  
Light means a noncohere amount of electromagnetic radiation, without 
coherence to other sources including matter so that only a quantum 
coherence between adjacent photons could matter\footnote{Unpublished.  
The quantization of the previous action and descendents 
are available upon request.}.  In 
regards to the wire configuration, special temporal retardation may 
be lossy due to unexpected light; in dark, light can be added uniformly 
with a flash synchronized with the current to enhance the bipolarimetric 
mingling of light and wire current with regards to the detector.  Light 
is bipolarimetric due to its form and content, and wire is too with 
regards to duration and timing of pulse when inundated by light; the final 
result is bipolarimetric lensing and possibly 
gravitational lensing in an experimental apparatus.  A laser diode of 
magnitude MeV will should mesh with the wire, given amgiguity in the 
uncalculated magnitute of the emitted light flash.  Numerical simulation 
is advised, and the macroscopic coherence of light could coher as 
a result long range in the tens of meters.   Numerical simulation is 
advised but due to the long range coherence and low intensity there is 
no more danger with the previous coherence than turning on the room light 
but focussed, which most likely has happened without measurable effect 
and is safe to try; neighboring beams at the CERN laboratory could get a 
noticeable gain in certain beams, even by sequencing them in a fraction 
of a second distinguished by the photoemitter type, and noticeable safety 
by not turning them on and off in a haphazard manner in the remaining beams 
with contiguity.

\vfill\break

\end{document}